\documentclass[12pt, a4paper, hidelinks]{article}

\usepackage{graphicx}
\usepackage{hyperref}
\usepackage{xcolor}
\usepackage[english]{babel}
\usepackage[nottoc]{tocbibind}
\usepackage[utf8]{inputenc}
\usepackage[T1]{fontenc}
\usepackage[backend=biber,style=alphabetic, defernumbers=true]{biblatex}
\usepackage[left=2.5cm,right=2.5cm,top=2.5cm,bottom=2.5cm]{geometry}
\usepackage{acronym}
\usepackage{fancyhdr}
\usepackage{lipsum}
\usepackage{array, tabularx, booktabs} 
\usepackage{longtable}
\usepackage{titlesec}
\usepackage{amssymb}
\usepackage{todonotes}
\usepackage{nicematrix}
\usepackage[autostyle=true]{csquotes}
\usepackage{cleveref}

\titleformat{\section}[hang]{\huge\bfseries}{\thesection\hspace{20pt}}{0pt}{\huge\bfseries}
\graphicspath{{figures/}{./}{assets/}}

\defineshorthand[english]{"~}{\babelhyphen{nobreak}}
\addto\extrasenglish{
  \languageshorthands{english}
  \useshorthands{"}
}

\newcommand{\thesistitle}{Chat AI: A Seamless Slurm-Native Solution for HPC-Based Services}

\newcommand{\authorname}{Ali Doosthosseini, Jonathan Decker,\\ Hendrik Nolte, Julian M. Kunkel}
\newcommand{\university}{GWDG\\Georg-August-Universität Göttingen}
\newcommand{\department}{Institute of Computer Science}
\newcommand{\thesistype}{Research Paper}

\newcommand{\keywords}{LLM, chatai, AI, hpc} 

\hypersetup{pdftitle=\thesistitle} 
\hypersetup{pdfauthor={Ali Doosthosseini, Jonathan Decker, Hendrik Nolte, Julian M. Kunkel}} 
\hypersetup{pdfkeywords=\keywords} 

\DeclareCiteCommand{\footcitetitleurl}[\mkbibfootnote]
  {\boolfalse{citetracker}%
   \boolfalse{pagetracker}%
   \usebibmacro{prenote}}
  {\ifciteindex
     {\indexfield{indextitle}}
     {}%
   \printfield[citetitle]{labeltitle}
   \setunit{\adddot\space}
   \usebibmacro{url+urldate}}
  {\multicitedelim}
  {\usebibmacro{postnote}}
  
\DeclareCiteCommand{\citetitleurl}[\mkbibbrackets]
  {\boolfalse{citetracker}%
   \boolfalse{pagetracker}%
   \usebibmacro{prenote}}
  {\ifciteindex
     {\indexfield{indextitle}}
     {}%
   \printfield[citetitle]{labeltitle}
   \setunit{\adddot\space}
   \usebibmacro{url+urldate}}
  {\multicitedelim}
  {\usebibmacro{postnote}}

\begin{document}

\fancyhead{}
\fancyhead[R]{\footnotesize \thesistitle}
\fancyfoot{}
\fancyfoot[R]{\thepage}
\fancyfoot[L]{Section \thesection}
\fancyfoot[C]{\authorname}
\renewcommand{\headrulewidth}{0.4pt}
\renewcommand{\footrulewidth}{0.4pt}

\pagestyle{plain}

\begin{titlepage}
\begin{minipage}[t]{0.6\textwidth}
\begin{flushleft}
\includegraphics[width=6.5cm]{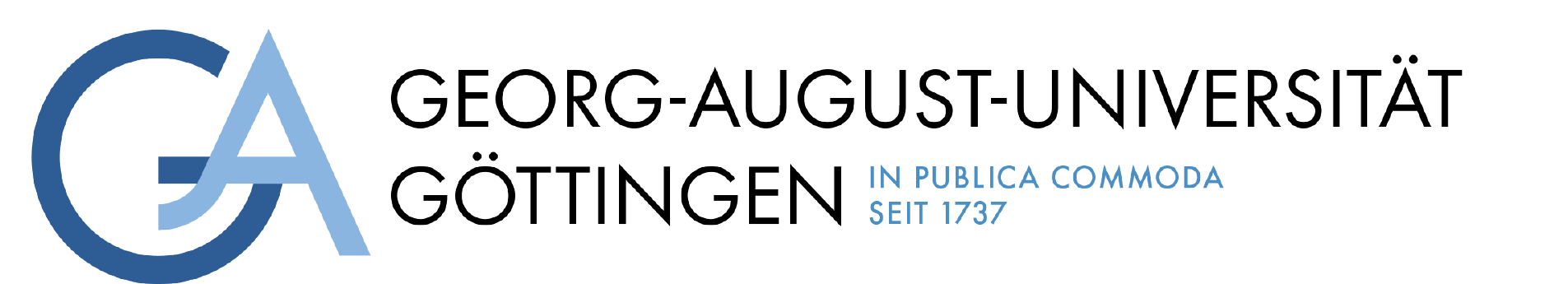}
\end{flushleft}
\end{minipage}
\begin{minipage}[t]{0.4\textwidth}
\begin{center}
\qquad\includegraphics[width=4cm]{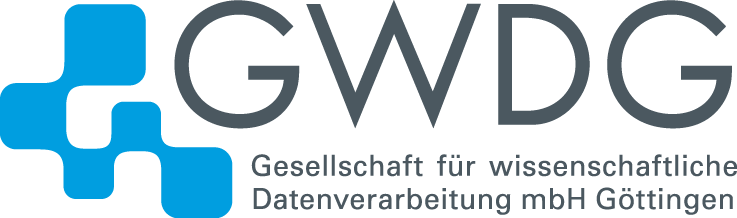}
\end{center}
\end{minipage}

\begin{center}

\vspace*{.06\textheight}
\LARGE \thesistype\\[0.5cm]

\rule{.9\linewidth}{.6pt} \\[0.4cm]
{\huge \bfseries \thesistitle}\vspace{0.4cm}
\rule{.9\linewidth}{.6pt} \\[1.5cm]

\Large\authorname\\
\hfill\\
\university\\
\department
\vfill
{\large \today}\\[4cm]

\vfill
\end{center}
\end{titlepage}


\newpage
\pagenumbering{roman}
\setcounter{page}{1}

\section*{Abstract}
The widespread adoption of large language models (LLMs) has created a pressing need for an efficient, secure and private serving infrastructure, which allows researchers to run open source or custom fine-tuned LLMs and ensures users that their data remains private and is not stored without their consent.
While high-performance computing (HPC) systems equipped with state-of-the-art GPUs are well-suited for training LLMs, their batch scheduling paradigm is not designed to support real-time serving of AI applications.
Cloud systems, on the other hand, are well suited for web services but commonly lack access to the computational power of HPC clusters, especially expensive and scarce high-end GPUs, which are required for optimal inference speed.
We propose an architecture with an implementation consisting of a web service that runs on a cloud VM with secure access to a scalable backend running a multitude of LLM models on HPC systems.
By offering a web service using our HPC infrastructure to host LLMs, we leverage the trusted environment of local universities and research centers to offer a private and secure alternative to commercial LLM services.
Our solution natively integrates with the HPC batch scheduler Slurm, enabling seamless deployment on HPC clusters, and is able to run side by side with regular Slurm workloads, while utilizing gaps in the schedule created by Slurm.
In order to ensure the security of the HPC system, we use the SSH ForceCommand directive to construct a robust circuit breaker, which prevents successful attacks on the web-facing server from affecting the cluster.
We have successfully deployed our system as a production service, and made the source code available at \url{https://github.com/gwdg/chat-ai}

\vfill

\newpage

\clearpage
\phantomsection\pdfbookmark{\contentsname}{toc}
\tableofcontents

\newpage
\clearpage\phantomsection
\listoftables
\phantomsection
\listoffigures

\thispagestyle{plain}
\newpage


\pagenumbering{arabic}
\setcounter{page}{1}
\pagestyle{fancy}

\section{Introduction}

Recent advances in large language models (LLMs)~\cite{brownLanguageModelsAre2020, openaiGPT4TechnicalReport2024, chowdheryPaLMScalingLanguage2022} have sparked an adoption of AI systems in a wide range of applications\footcitetitleurl{parmarCouncilPostGenerative}~\cite{openaiGPT4TechnicalReport2024}.
This puts many institutions and companies in the difficult position of having to come up with a strategy to adopt these new technologies and deciding to what extent their data can be handled by (potentially foreign) companies such as OpenAI and Google~\cite{openaiGPT4TechnicalReport2024, chowdheryPaLMScalingLanguage2022} or alternatively attempt to self-host an LLM service in order to preserve the privacy of user data~\cite{sebastianPrivacyDataProtection2023, kshetriCybercrimePrivacyThreats2023}.

High-quality LLM models have recently become widely available due to many companies and institutions providing free access to their model weights, commonly via Hugging Face\footcitetitleurl{HuggingFaceAI2024}.
Notable recent releases include Llama3.1 via Meta~\footcitetitleurl{LlamaHerdModels}, Qwen2 via Alibaba Cloud~\cite{qwen2} and Mixtral via Mistral AI~\cite{jiangMistral7B2023}.

Using a runtime such as llama.cpp\footcitetitleurl{GgerganovLlamaCpp}, vLLM~\cite{10.1145/3600006.3613165} or Text-Generation-Inference\footcitetitleurl{HuggingfaceTextgenerationinference2024} one can run smaller versions of these models, e.g., 7B or 8B parameters, on high-end consumer GPUs with 10 to 20\,GB of VRAM.
However, for larger models such as Llama3.1-70B~\footcitetitleurl{MetallamaMetaLlama3170BInstruct2024} and Qwen2-72B~\cite{qwen2} that more closely rival the performance of state-of-the-art closed source models such as OpenAI's ChatGPT4~\cite{openaiGPT4TechnicalReport2024}, significant investment into GPU resources is required.

In our testing we found that a single instance of Meta's LLama3.1-70B model fits into the VRAM of two of NVIDIA's H100 GPUs when using FP8~quantization~\cite{wuZeroQuantFPLeapForward2023, micikeviciusFP8FormatsDeep2022} via vLLM~0.5.0 or even a single H100 GPU when significantly reducing the context size.
While this allows effective inference of LLama3.1-70B and models of similar size, it limits access to private inference of these models to those who can afford H100 GPUs or similarly powerful hardware.

Many research centers, universities and institutions either operate their own HPC infrastructure or have access to HPC infrastructure via a research network\footcitetitleurl{OurMembersNHRVerein, Members}.
Typically, HPC centers operate a batch scheduler such as Slurm~\cite{yooSLURMSimpleLinux2003} and put strong emphasis on the security of their systems.

HPC infrastructure is an attractive target for attackers to steal research data, interrupt operability or hijack compute resources for crypto mining.
To ensure system security, on HPC multi-user systems, admins often rely on battle-tested security mechanisms such as POSIX permissions and SSH for access management.
This significantly restricts the possibility of using HPC resources to expose services without compromising these security frameworks.

Furthermore, a batch scheduler such as Slurm requires users to submit jobs, which it then schedules at a suitable time to one or more compute nodes depending on the expected job duration, resource demands and availability of resources.
Considering that a user of an interactive HPC service, e.g., chatting with an LLM, would expect their request to be processed immediately without waiting for a Slurm job to be scheduled and potentially waiting on other jobs, the batch scheduling paradigm is not suitable for offering services.

In this paper, we present an entirely Slurm-native solution that seamlessly integrates with existing HPC infrastructures to offer a demanding service such as real-time conversations with LLMs with minimal effort.
Moreover, as it is Slurm-native, it allows for the effective utilization of HPC resources while not permanently blocking resources for other Slurm jobs.
Using this method, we released Chat AI\footcitetitleurl{ChatAI}, a pioneering web service that offers rapid real-time access to state-of-the-art open source LLMs hosted on our existing HPC infrastructure.
This solution addresses the need for Slurm-based services, private inference via local or associated HPC centers and securing said services using SSH-based access patterns that comply with many existing HPC security frameworks.
We share our findings and explore the viability of hosting a service with the proposed architecture with regards to security, performance, and user adoption.

This paper is structured as follows:
\Cref{sec:background} provides the necessary background information to understand the rest of the paper.
\Cref{sec:relatedwork} places our work in the context of existing publications and explores similar approaches.
The significant challenges we faced during the design and implementation of our solution are discussed in \Cref{sec:challenges}.
Our architecture is then presented in detail in \Cref{sec:architecture}.
Next, we evaluate our service in terms of security and performance in \Cref{sec:evaluation}, and discuss its adoption by users and institutions, as well as its limitations and future prospects in \Cref{sec:discussion}.
Finally, we summarize our contributions in \Cref{sec:conclusion} and provide our source code and acknowledgements in \Cref{sec:code}.

\section{Background}\label{sec:background}

In this section we describe the existing tools and paradigms that are relevant to our approach including
\begin{itemize}
    \item the HPC batch scheduler Slurm,
    \item the scheduling paradigms for batch and service scheduling,
    \item the open source LLM runtime vLLM,
    \item the API gateway Kong and
    \item some notes on security aspects for HPC infrastructure.
\end{itemize}

\paragraph{Slurm} The Simple Linux Utility for Resource Management, in short Slurm~\cite{yooSLURMSimpleLinux2003}, is a workload or resource manager for HPC clusters.
It offers a solution for running batch jobs on a cluster through the use of shell scripts.
Slurm assigns jobs to nodes depending on the availability of the current nodes and requirements of the job, ensuring that the resources of the cluster are utilized efficiently.

\paragraph{Scheduling Paradigm}
We consider two major scheduling paradigms, batch scheduling, also called gang scheduling, and service scheduling.
In batch scheduling, when assigning a job that requests a number of resources such as multiple nodes, a batch scheduler such as Slurm will only assign and start a job when all resources are available at once for the requested duration.
While this is desirable for a compute job such as an MPI job that can only start when all nodes are available to use batch scheduling, it is a different case for services.
For a service that is supposed to serve user requests, multiple instances might be required to sufficiently serve all requests but if not enough resources are available, the service could still operate with reduced performance and dynamically add instances when more resources become available.
For one-time jobs that are not time-sensitive, Slurm is a suitable choice, however, one would have to adapt it in order to support a service paradigm.

\paragraph{vLLM}

vLLM~\cite{10.1145/3600006.3613165} is a state-of-the-art, open source LLM runtime implementation.
It uses an attention mechanism inspired by memory paging techniques in operating systems, achieving near-zero waste in key-value cache memory.
Since its first release in June 2023\footcitetitleurl{VllmprojectVllm2024} vLLM has kept up with many developments and optimizations for LLM inference techniques with the latest version, as of writing, v0.5.0.post1, having been released on Github in June 2024.
vLLM supports modern LLM architectures such as Llama and Qwen and implements an OpenAI-compatible API, such that it is a drop-in replacement for most applications with ChatGPT integration.

\paragraph{Kong}

Kong\footcitetitleurl{BecomeAPIfirstCompany} is an API gateway that supports custom plugins, complex routing, user management and monitoring tools.
It consists of an open source edition, Kong OSS\footcitetitleurl{KongKong2024} and an enterprise edition with additional features.
Here we focus on the open source edition, which supports creating API routes and keys, implements native user management and provides features such as rate-limiting and restricting access for certain routes and users.

\paragraph{Security Aspect}
Security is a major concern when thinking about hosting a service that is exposed to the internet on an HPC cluster as this opens a much wider attack surface for attackers compared to regular access patterns via SSH~\cite{bulusuAddressingSecurityAspects2018}.
If an attacker should be able to break into the system, they would be able to steal potentially sensitive research data, perform Distributed Denial of Services (DDoS) attacks on other systems~\cite{burgerIntroductionWebService2012} or run crypto-miners on the HPC hardware~\cite{pottOvercomingPitfallsHPCbased2023}.
To protect themselves, HPC centers commonly employ strict security regulations such as disallowing users access to root permissions on compute and login nodes, limiting the allowed access protocols, e.g., only SSH, and restricting compute nodes to not have direct internet access with only cluster login nodes being reachable from the internet.

\section{Related Work}\label{sec:relatedwork}

Providing access to LLMs is in itself a service offering, and running such services in a production environment is commonly associated with existing frameworks such as Kubernetes~\cite{liuConvergenceHighPerformance2023}.
However, most HPC infrastructures rely on batch scheduling systems such as Slurm, Torque or HTCondor~\cite{liuConvergenceHighPerformance2023, hildrethLargescaleHPCDeployment2020} as their main resource manager.
Moreover, modern LLMs require so much compute power not only for training but also for inference~\cite{wangEfficientReliableLLM2024, wangGPUAcceleratedMachineLearning2021} that an inference service would benefit from access to powerful hardware such as high-end GPUs installed in HPC infrastructure.

In order to use these HPC resources, various approaches exist.
\begin{itemize}
    \item Zhang et al.~\cite{zhangArtificialIntelligencePlatform2019} proposed a unified framework for Apache Hadoop YARN~\footcitetitleurl{ApacheHadoopApache} and Slurm to create a mobile AI service platform.
    \item Wan et al.~\cite{wanOpenVenusOpenService2023} proposed a Slurm-based service platform to optimize startup times and storage usage for multi-user applications on HPC infrastructure.
    \item Shainer et al.~\cite{shainerNVIDIACloudNative2022} proposed a service architecture with a focus on supporting specialized HPC hardware such as DPUs and Infiniband.
    \item Zhou et al.~\cite{zhouContainerOrchestrationHPC2021b} proposed an architecture that utilizes Torque to run compute jobs submitted via an interface running on a Kubernetes cluster.
    \item Metje~\cite{metjeRunningKubernetesWorkloads2024} proposed an approach and studied various existing approaches for a running Kubernetes cluster on top of Slurm without relying on root access.
\end{itemize}

On a given HPC system using Slurm as its resource manager, trying to employ some of these approaches would require replacing Slurm with another scheduling system such as Kubernetes.
This would disrupt existing users and workflows of such a system and force users to migrate to the new system.
By keeping the existing resource manager, the inference service could operate side by side with user workloads and utilize gaps in the Slurm scheduling~\cite{przybylskiUsingUnusedNonInvasive, copikSoftwareResourceDisaggregation2024}.
While this approach has the potential to greatly improve the resource utilization of the HPC hardware, for example, by serving inference from gaps in the scheduling besides training jobs, this can also result in a reduced user experience for users of the inference service.
Especially when serving large LLMs, long start-up times can be expected when loading these models into memory.
Where serverless functions commonly take only seconds to load~\cite{przybylskiUsingUnusedNonInvasive}, loading an LLM can take multiple minutes, which would be the time a user needs to wait for a response to their inference request if no instance of the desired model was ready.
This issue becomes especially pressing when considering the high need for GPU resources from LLM inference and training workloads, which could result in training jobs consuming all GPUs in a cluster and leaving no resources for inference services or vice versa.

Another major aspect of running services on HPC systems is security~\cite{brayfordDeployingAIFrameworks2019} as sensitive research data and valuable compute resources are concentrated in HPC centers.
A common access protocol for HPC clusters is SSH~\cite{liuConvergenceHighPerformance2023}.
SSH itself allows for creating a remote terminal but can also be used to forward arbitrary ports or commands to allow remote access to web servers~\cite{BURGER2012334}.
Using SSH to tunnel traffic to create a web service with a secure connection to a backend running on HPC infrastructure is an approach for creating a HPC-based inference platform, but storing an SSH key in an exposed server could pose a security risk and incentivize bad actors to penetrate the server in order to gain access to the HPC cluster, and may thus not be allowed in HPC centers with strict security standards such as SSH access only from individual users on local devices.

One should also consider the possibilities for the usage of LLMs in education~\cite{lanTeachersAgencyEra2024, jeonLargeLanguageModels2023} as many HPC centers are closely connected to universities and could offer inference capabilities to students as an alternative to commercial offerings.

Besides the efforts described in this paper towards a private LLM inference, other research centers also made efforts to develop a private chat interface including FhGenie~\cite{weberFhGenieCustomConfidentialitypreserving2024a}, HAWKI\footcitetitleurl{environmentsHAWKDigitalEnvironmentsHAWKI2024} and Minverva Messenger\footcitetitleurl{MinervaMessenger}.
Their approaches to privacy are mainly to implement user management on their side and to use a single OpenAI API key to send all requests to the external ChatGPT API in order to obfuscate the attribution of a given request to a specific user, essentially acting as a wrapper or middleman for ChatGPT.
The privacy provided by this approach is, however, limited as all requests and responses are still sent to a third party, e.g., OpenAI or Microsoft Azure.
Support for alternative endpoints via self-hosted backends that provide an OpenAI-compatible API, which is in development for HAWKI and FhGenie, could provide a fully private solution.

\section{Challenges}\label{sec:challenges}

In this section we present the challenges and concerns of hosting web services on an HPC infrastructure that employs Slurm as its resource manager.
As challenges we consider
\begin{itemize}
    \item paradigm differences of using a batch scheduler, such as Slurm, to provide a service,
    \item security implications when exposing a web service connected to HPC infrastructure,
    \item performance inhibitors that a service running on HPC faces, 
    \item dynamic scalability when using a batch scheduler,
    \item reliability features, such as health checks and self-healing,
    \item the options for guaranteeing user data privacy and
    \item monitoring and accounting.
\end{itemize}
Each of these points is discussed in detail in the following.

\begin{paragraph}{Paradigm Differences}
The purpose of HPC clusters is commonly to run large-scale and demanding batch jobs with a fixed start and end point, but not continuous and scalable web services.
The batch job and service paradigms are strictly different from one another and have different expectations and goals.
Slurm is designed to handle the batch job paradigm and will schedule incoming jobs according to it.
In order to adapt it to the service paradigm, a job that runs an instance of a service must continuously be replaced or extended.
This requires some other scheduling control flow to operate on top of Slurm to implement this functionality, e.g., keeping a pool of one or more instances active for a given service by ensuring that one or more jobs are active for that service at a given time.
\end{paragraph}

\begin{paragraph}{Security}
From a security perspective, hosting a web service on an HPC cluster opens up additional attack surfaces.
An HPC cluster is a valuable target for an attacker as it offers massive amounts of compute resources, which could be misappropriated for crypto mining and are hosts for potentially sensitive research data.

A web service that is intended to be reachable via the public internet requires some components of that service to be exposed from behind the firewalls of a given HPC center.
This would open an attack surface for malicious actors to gain a foothold on the HPC infrastructure by exploiting a vulnerability in the web service.
While the implementation of a given web service should avoid vulnerabilities, a defense-in-depth approach that prevents further exploitation, is required.

SSH is an access protocol commonly allowed on HPC infrastructure and in some cases is the sole option to access the necessary HPC components, thereby forcing access patterns to build on top of SSH to remain compatible with the existing security framework.
Moreover, as an SSH connection is established using cryptographic keys, which would need to be stored on a potentially exposed machine in order to connect to the HPC backend, extra caution must be taken to ensure the damage caused by a potential compromise of this key is limited, such that it cannot be used as an entry point to the HPC cluster and its compute power by an attacker.
\end{paragraph}

\begin{paragraph}{Performance}
HPC infrastructure commonly includes specialized hardware such as HPC interconnects, e.g., Infiniband, and parallel file systems to optimize the performance of batch jobs.
While HPC hardware and Ethernet interconnects are normally powerful enough to support most services, when pushing the performance, any introduced workarounds may result in bottlenecks and add latency or hinder throughput.
An infrastructure that is dedicated and tailored to running production services may therefore consistently outperform the architecture proposed in this paper.
\end{paragraph}

\begin{paragraph}{Scalability}
Slurm does not inherently provide an option to scale up the number of jobs as the number of requests increases, requiring additional efforts to implement automatic scaling.
In order to support a service where request volumes vary over time, it is necessary to support automatic scaling, where the number of instances for a given service is adjusted based on request volume.
This requires the implementation of a functionality to measure the current request volume as well as to calculate and enforce the desired amount of service jobs.
Moreover, this introduces the problem of service discovery, as incoming requests need to be load balanced across all available service job instances and therefore the system responsible for routing incoming requests needs to be aware of all healthy job instances.
\end{paragraph}

\begin{paragraph}{Reliability}
Slurm itself is robust against nodes failing by ignoring problematic nodes and assigning jobs to other nodes until an administrator fixes the issue.
However, Slurm does not implement mechanisms to ensure constant availability such as health checks, which are crucial for a service to be reliably accessible and available at all times.
These mechanisms therefore need to be implemented on top of Slurm.
\end{paragraph}

\begin{paragraph}{Privacy}
For a service that offers an interface to chat with LLMs, essentially receiving user inputs and responding with outputs generated by LLMs that process these conversations, one can consider all user inputs and responses to be sensitive user data that must be kept private, thereby limiting the options for storing user data.
Some approaches include using cryptography and key management mechanisms to keep user data encrypted and inaccessible while it is not used directly.

However, in the context of highly sensitive information such as medical data, privacy is critical and should be prioritized, as leakage of such data could be catastrophic.
Depending on the type of service, stricter privacy requirements can be achieved by enforcing the use of encryption keys for users to be able to download and decrypt their data.
The other option, which is feasible and arguably even more secure, is to never store sensitive user data, such as past and ongoing conversations with LLMs, on the server side and instead store all such data only on the local devices, for example in the browser.
Another consideration to data privacy is the encryption of data in transit in order to prevent a scenario where an attacker performs man-in-the-middle attacks after compromising part of the service to read user inputs and responses from the services.

It is possible to implement end-to-end encryption and to keep data encrypted during computation using Trusted Execution Environments, which are currently available for some NVIDIA GPUs\footcitetitleurl{DataIntegrityAI}.
These additional measurements can be considered for system hardening but go beyond our basic requirements, as these serve to protect incoming data while parts of the system have already been compromised.
Nevertheless, when following defense-in-depth security paradigms, these additional measurements are well worth pursuing.
\end{paragraph}

\begin{paragraph}{Monitoring and Accounting}
When submitting jobs via Slurm, it is possible to perform accounting based on the compute hours used by a given job and attribute them to the user who submitted the job.
However, when offering a service, it is to be expected that multiple users will be served by the same job instance.

Furthermore, it can also be expected that many service users do not have access privileges to the HPC systems, such that the service user management is independent from the HPC center.
Therefore, the service jobs should be submitted via a functional account that has no correspondence to an individual user or administrator.
User-based monitoring, accounting and moderation instead has to be handled in an authentication layer in front of the service or integrated within the service itself.
Individual requests can be attributed to logged-in user accounts, and the timestamp for a given request can be stored in order to monitor the service usage without violating the privacy assumptions from the previous paragraph.
\end{paragraph}

\section{Chat AI Architecture}\label{sec:architecture}

\begin{figure}[ht]
    \centering
    \includegraphics[width=0.8\textwidth]{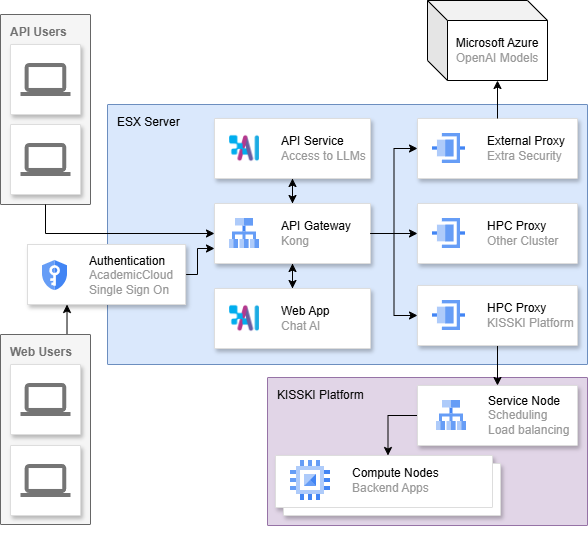}
    \caption[Architecture of Chat AI.]{Architecture of Chat AI. This diagram displays the main components of the service, consisting of an ESX web server that communicates to the login/service node, and the compute nodes of the HPC KISSKI platform.}
    \label{fig:arch-all}
\end{figure}

The architecture of Chat AI is shown in \Cref{fig:arch-all} with its two major components, namely the web server labeled as \textit{ESX Server} and the HPC infrastructure labeled as \textit{KISSKI Platform}.
Fundamentally, the HPC platform is only expected to provide Slurm itself.

Each of the two major components is split up into further sub-components, which are individually explained in the following.
It should also be noted that all sub-components running on the ESX Server have been containerized and operate via docker compose.

\subsection{Authentication}
We employ an Apache web server as a reverse proxy for our web interface to provide integration with our Single Sign-On (SSO) provider Academic Cloud\footcitetitleurl{AcademicCloud}.
This is effectively an OAuth2 authentication mechanism using the Apache web server module OpenIDC and could be used via any other SSO provider that implements the same protocols.

Once a user has signed into the SSO provider, they are forwarded through the API Gateway to the Chat AI web interface.
Furthermore, when forwarding the requests, the Apache web server attaches the email address of a given user as the user id to all requests.

\subsection{API Gateway}

We deployed the Kong Open Source edition\footcitetitleurl{KongKong2024} as the API gateway to route incoming requests to the corresponding upstreams, e.g., the HPC proxy or the web app.
Kong was chosen as it is open source and provides all the features we require from an API gateway including creation and management of API routes, rate limiting, API key management and observability.

Continuing from the Apache web server authentication component, a user request is forwarded by the Kong API Gateway to the Chat AI web interface via its respective route.
If a user then submits an inference request to an LLM via the web interface, the request is sent through the API Gateway to the corresponding route depending on the selected model.

Moreover, the API Gateway also exposes an access point for API users.
Users with valid API keys can bypass the Apache web server and connect to the API Gateway directly, which performs authentication via the provided API key.
This unifies the path for web and API users past the API gateway such that the incoming route and user is abstracted from the backend service.

\subsection{Web App}

\begin{figure}[ht]
    \centering
    \includegraphics[width=0.8\textwidth]{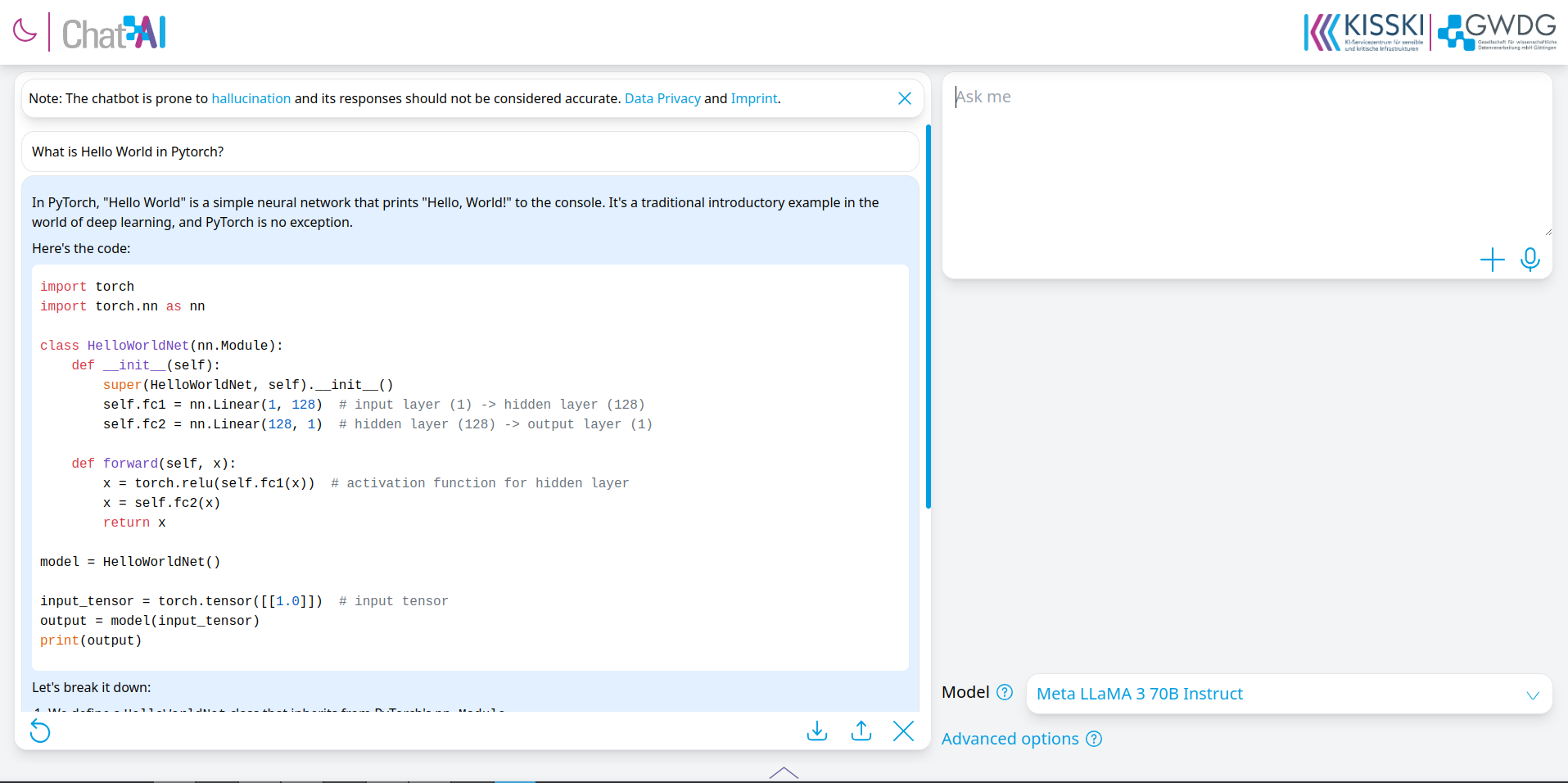}
    \caption[Chat AI App.]{Chat AI App. This shows the Chat AI web interface written with React and Vite with the chat history on the left, the prompt window on the top right and a drop down for model selection at the bottom right.}
    \label{fig:web}
\end{figure}

The web interface illustrated in \Cref{fig:web} serves as a user-friendly way for users to interact with a selection of LLMs.
After testing existing open source software such as Text-Generation-Webui\footcitetitleurl{oobaboogaOobaboogaTextgenerationwebui2024} or LibreChat\footcitetitleurl{avilaDannyavilaLibreChat2024} we found that we require more tight control over the handling of user data and the routing of requests, which we gained by implementing our own web interface.
Most notably we wanted to ensure that our application exclusively runs in the browser with no significant active server code, which prevented us from employing a solution such as HAWKI\footcitetitleurl{environmentsHAWKDigitalEnvironmentsHAWKI2024} that relies largely on server-sided execution via PHP.

We experimented with various solutions and built a prototype for the web interface using gradio~\cite{abidGradioHasslefreeSharing2019} but this resulted in significant load on our server for large user numbers.
Afterward, we tried  to rewrite the web interface using React\footcitetitleurl{React} and Vite\footcitetitleurl{Vite}, which is the design that is shown in \Cref{fig:web}.
With this implementation we achieved a significant load reduction on our web server.

In \Cref{fig:arch-all} the Chat AI web app is shown behind both the authentication and the API gateway from the user perspective as only the API gateway is exposed externally and the web app itself is served as a route via the API gateway.

\subsection{HPC Proxy}

To enable the submission of inference requests from the web server to the HPC cluster without compromising its tight security, we wrap this communication in an SSH connection.
SSH is the only protocol allowed to access the login node of our HPC center, labelled as \textit{Service Node} in \Cref{fig:arch-all}, and automatic logins via SSH to user accounts are also forbidden.

We deploy an SSH key on the web server and prevent it from potentially being used by an attacker to gain a foothold on the HPC cluster by employing the ForceCommand directive of SSH\footcitetitleurl{Sshd_configLinuxManual}.
This directive is used to restrict an SSH key to a specific command such that it cannot be used to run generic shell commands on the remote machine. 
Moreover, as this is configured via the \textit{authorized\_keys} file in a user's directory, it can be configured without root access.
In our case, the SSH key belongs to a functional account which is also responsible for the submission of service jobs to Slurm.

We have developed the HPC Proxy as an application that ensures that the SSH connection between itself and the HPC login node remains open and is reestablished in case an interruption breaks the SSH connection.
Interruptions are detected by attempting to send keep alive pings once every 5 seconds.
Besides that, the HPC Proxy serves as a regular proxy that forwards requests to our Cloud Interface Script, a bash script running on the HPC service node.
The Cloud Interface Script is designed to forward authorized, inference-related HTTP requests to a service node, which is running the desired model and ready to respond.
Additional steps were taken to ensure the proxy is fast, efficient and generalizable with support for various types of requests and responses, including streaming.

In \Cref{fig:arch-all} we included two HPC Proxy instances, as this architecture decouples the web server from the HPC platform, allowing a single web server to potentially utilize multiple HPC platforms by starting an HPC Proxy instance per HPC platform and load balancing via the API Gateway.
On the other hand, it is also possible to have multiple web servers utilize the same HPC platform, each with its own SSH connection.
With this configuration, we are able to provide custom web services that require HPC resources without directly exposing the components in the cluster.

\subsection{Cloud Interface Script}

This is bash script is placed on the service node and receives every request send via the HPC Proxy.
The ForceCommand directive for the SSH connection is configured such that the HPC Proxy is only able to call the Cloud Interface Script via the SSH connection.
On every keep alive ping, set to every 5 seconds for the SSH connection, it triggers our scheduler script, which is discussed in the next section.

The HPC Proxy transmits HTTP requests in the form of parameters when calling the cloud interface script via an SSH command and receives the responses via stdout from the cloud interface script.
For larger requests, the cloud interface script receives the requests via stdin.
The cloud interface script utilizes a routing table maintained by the scheduler script to determine where to forward the request.

\subsection{Scheduler}

We developed a scheduler script, which is run periodically on every incoming keep alive ping from the HPC Proxy, with the goal of ensuring the availability of the service instances.
The scheduler script can be configured with a set of services it should maintain along with the specifics of running their respective jobs, such as the job script and settings for when to adjust the number of active instances for a given model.

The scheduler script obtains a list of running jobs for each service from Slurm via the \textit{squeue} command, compares it against the given configuration, and takes the necessary actions to maintain the services, such as submitting new jobs via \textit{sbatch}.
In order to avoid race conditions, we ensure only one instance of the scheduler is running at a time by means of a lock file.

The scheduler script maintains a routing table, containing an entry for each active job, along with its associated service, node and port number.
This table is used to route each incoming request to a specific node and port number associated with a service instance, selected randomly out of all eligible instances, effectively performing random load balancing.

As Slurm provides no network virtualization, if two jobs on the same node try to occupy the same port, it will fail for the second job as the port is already occupied by the first job.
To avoid this, the scheduler script picks a random port when submitting a new job for that job to use and checks the routing table to ensure that the port is not already occupied.

Once a service job has been submitted, it takes an uncertain amount of time before the instance is ready to serve requests, as Slurm needs to assign it to a node and the model needs to be loaded into GPU memory.
Therefore, the scheduler script periodically probes the newly submitted jobs until they are ready, before marking them as ready to serve requests in the routing table.

To scale up and down the number of service instances according to demand, calculating the request volume for a given service is necessary.
This is performed by storing the average number of concurrent requests to each service within a predefined time window, which is then recalculated and updated on each scheduling run.
If this average is higher than a certain threshold, the scheduler spawns multiple instances of that service to prevent the formation of a backlog.
Likewise, when the average is too low, the scheduler allows the excess jobs to expire without resubmitting them, effectively reducing the number of instances and freeing up resources.

In an alternative design, the request volume could be measured by the API Gateway or HPC Proxy and forwarded to the scheduler script, but we decided against this option in order to minimize the coupling between the web server and the HPC platform, and instead pursued a solution to calculate the request volume directly on the HPC platform.

\subsection{LLM Server}

As Chat AI is an LLM chat service, we also require a runtime to serve LLM inference requests.
For that purpose, one can assume the individual services mentioned in the previous subsection to refer to various LLMs hosted on the HPC infrastructure, but they can also be any server runtime that benefits from HPC resources.

To run an LLM model with GPU acceleration, we employ the open source runtime vLLM\footcitetitleurl{VllmprojectVllm2024}.
vLLM has the advantages of offering an OpenAI-compatible API and an active community, which quickly implements and supports many popular open source LLMs.
From our experience, vLLM was several times more efficient than our unoptimized LLM runtime implementation.

\subsection{External Proxy}

The External Proxy serves as an optional extension to the architecture to provide access to external models such as ChatGPT4~\cite{openaiGPT4TechnicalReport2024} as an additional route in the API Gateway, effectively becoming a wrapper for additional services.
As ChatGPT4 requires paid access, we placed strict rate limits via the API Gateway and restricted access to certain user groups.

\subsection{Monitoring}

The monitoring components are not shown on \Cref{fig:arch-all} as they are optional but nevertheless important for capturing metrics and detecting issues.
We employ an external Grafana service, which constantly captures logging and monitoring data from a Prometheus server that is integrated into the API gateway via a plugin.
This is performed via standardized monitoring endpoints and ensures only authorized access to the monitoring system.

\section{Evaluation}\label{sec:evaluation}

Our primary goals for the LLM web service were to ensure its privacy, security and practicality.
Due to the nature of our infrastructure and demands of our users, security and privacy took the highest priority in our design.
In this section, we evaluate the architecture for Chat AI based on the achieved standards for security and data privacy as well as performance measurements regarding throughput and latency.
Furthermore, we assess the user adoption as a factor in determining the architecture's real-world viability and overall success.

\subsection{Security}
\label{subsec:security}
In order to evaluate the security of our architecture we assume different attack scenarios.
These should serve to visualize that our designs follow the principals of defense-in-depth, minimizing permission and not storing anything that does not need to be stored.

\subsubsection{Security of the Web Interface}
Our Chat AI web interface employs standard security mechanism via TLS and in order to access it, a user must have a valid account and login to authenticate with the SSO provider.

Assuming a breach in the web interface, which gives an attacker a shell in the web interface container, this would allow the attacker to spy on users of the web interface and set up man-in-the-middle attacks until the breach is discovered and patched.
However, as no user data is stored, no accumulated data can be stolen by an attacker.

Administrative tasks on the web server are handled via an SSH connection, which is only possible from our internal VPN when utilizing two-factor authentication.

Due to the containerized nature of the setup, an attacker would need to escape the container to gain a shell on the host system and then gain root access in order to compromise other parts of the system.
Even in this scenario, an attacker would not be able to use the compromised web server as an entry point into other parts of the HPC infrastructure.

\subsubsection{Security of the HPC System}
In our view, compromising the HPC system would be the worst case scenario that needs to be prevented at all costs.
As users do not have direct access to the HPC infrastructure in this architecture, any attacks on the HPC infrastructure would necessarily originate from the web server, meaning that some component within the web server would have to be already compromised.
In this case an attacker may attempt to locate and read the SSH key used to establish the connection to the HPC system.
However, as the SSH key is configured with ForceCommand to only ever call the Cloud Interface Script and corresponds to a functional account with no administrative permissions on the cluster, its effectiveness in penetrating the HPC platform is significantly reduced.

A potential alternative to the attack scenario above would be if an attacker discovered a flaw in our Cloud Interface Script and attempts to perform injection attacks by submitting requests that instead of being routed to a service, end up executing shell commands.
For this purpose we bring extra attention to the implementation of the input parsing in the Cloud Interface Script to protect against injection attacks, restricting any request to follow a preset of determined paths, and avoiding any potentially dangerous commands such as \textit{eval}.

Essentially, an attacker would need to breach multiple layers of security to first breach the the web server, and then penetrate multiple security layers again to get shell access on the HPC system.
Even if these security layers fail, the previous conversations and messages of our users physically cannot be stolen from us as they were never stored in the first place.
While it is impossible to claim that a system's security is perfect, we consider having multiple layers that an attacker would need to breach as a marker of our architecture's security.

\subsection{Data Privacy}
As for data privacy, we regard compliance with GDPR and the strict data privacy measures it necessitates as our main benchmark.
Sensitive and personalized data must be transmitted and stored securely to ensure that third parties cannot access it.
We take privacy a step further and minimized the data collection according to GDPR article~5(1)~lit.~c to the degree that no user prompts are being stored at any circumstance, i.e., prompts and responses are not stored on the server.
To still provide a stateful conversation, the conversation history is solely stored within the user's browser, which can be deleted with a button provided in the web interface or by resetting the browser cache. 

It follows that conversational data is also not used for any purpose such as training models.
Some non-conversational usage statistics are gathered solely for monitoring and accounting purposes.
This includes the user's account identifier, timestamps for requests, and selected model.
With this information, we can monitor the load on the individual models and identify potential misbehaving users.

As a result of these measures, an attacker on the web interface or on the HPC system cannot access past conversations as they are only kept locally on the users' devices and never stored on the server.

\subsection{Performance}

For our architecture to be able to support increasing numbers of users, we need to assess its performance.
To evaluate the effectiveness of incorporating the HPC cluster as the backbone of our service, we conducted a series of tests. 
These tests measured the throughput and latency of each component in the architecture and can be used to identify bottlenecks in the pipeline.
Latency measurements were performed using a custom shell script, and for throughput measurements we employed the load testing framework Locust\footcitetitleurl{LocustioLocust2024}.

\subsubsection{Latency}

The latency or response time is the amount of time it takes for a client to receive a response to a request, i.e., the duration between the time a request is sent and the point in time when the response is received.
To identify potential bottlenecks, we performed latency measurements for every individual component in the architecture on the path for user requests and calculated the time spent in each step.
For statistical validity, we took the average over 50 measurements with identical conditions.

The setup that the tests were performed on consists of a VM web server running Ubuntu 22.04.4 LTS with a 16-core AMD EPYC processor and 8\,GB of RAM, and a Slurm-based HPC cluster consisting of one login node and 10 GPU nodes, each with 4 Nvidia H100/80\,GB GPUs and a 52-core Intel(R) Xeon(R) Platinum 8470 processor with 500\,GB of RAM running Rocky~Linux~9.2.

\begin{table}[ht]
\caption[Latency measurements from the ESX machine.]{Latency measurements from the ESX machine.
Aggregated average latency for a request from the ESX machine to first token from LLM in active service job.
Probe refers to HTTP request on health endpoint.
Column 3 gives the aggregated average latency in ms with the difference, e.g., latency incurred in the respective operation.}
\centering
\begin{NiceTabular}{llrr}
\CodeBefore
\rowcolors{2}{white}{gray!10}
\Body
\textbf{Component} & \textbf{Operation} & \textbf{Agg. Avg. (std.) in ms} & \textbf{Diff. in ms} \\
 ESX Machine & Probe local proxy &  2.59 (0.56) & 2.59 \\
 HPC Service Node & SSH Command & 13.12 (0.59) & 10.54 \\
 HPC Service Node & Probe GPU node & 18.43 (1.86) & 5.30 \\
 HPC GPU Node & LLM First Token &  51.06 (2.03) & 32.63 \\
\end{NiceTabular}
\end{table}

Under normal conditions when the system is not overloaded, a user can expect to receive their first response token after approximately \(50\)\,ms, of which more than \(27\)\,ms are the compute time of the underlying LLM.
Therefore, the additional latency overhead introduced by our architecture was approximately \(23\)\,ms, which is not noticeable in most cases and we consider acceptable.

\subsubsection{Throughput}

\begin{table}[ht]
\caption[Throughput results for a regular user request.]{Throughput results for a regular user request.
\(^*\)The actual throughput was higher than the capacity of Locust to spawn requests in our setup, and was deemed sufficient for our purposes.
\(^{**}\)The prompt was \textit{\enquote{count from 1 to 10}}, to which the model almost always responded with the numbers 1 to 10 consecutively.}
\label{tab:throughput}
\centering
\begin{NiceTabular}{lr}
\CodeBefore
\rowcolors{2}{white}{gray!10}
\Body
 \textbf{Component/Operation} & \textbf{Throughput (RPS)} \\
 Apache Web Server& \(3\,000+^*\) \\
 Kong API Gateway & \(3\,000+^*\) \\
 Chat AI Web Interface &  \(1\,300\,-\,1\,800\) \\
 Chat AI Web Interface Middleware &  \(200\,-\,300\) \\
 SSH to HPC Service node &  \(200\) \\
 SSH to HPC GPU node &  \(200\) \\
 Single word from 7B LLM &  \(~100\) \\
 Sentence from Intel Neural 7B LLM\(^{**}\)& 27 \\
 Sentence from Mixtral 8x7B LLM\(^{**}\) & 8 \\
 Sentence from Qwen1.5 72B LLM\(^{**}\)  & 2 \\
 Sentence from Meta Llama3 70B LLM\(^{**}\)  & 2 \\
\end{NiceTabular}
\end{table}

The throughput or load experiments were conducted to assess the capacity of our architecture, i.e., its capability to handle a large number of requests from users.
This enables us to assess the efficiency and scalability of our architecture and its implementation.
Similar to the latency measurements, we individually tested the throughput for each of the individual components and steps in the architecture on the path of a user request.

The hardware setup was identical to that of the latency tests, but we used Locust for the throughput measurements.
The results are captured in \Cref{tab:throughput}, which shows the throughput in requests per second of every individual component of our infrastructure.

The selection of models used in this test include Intel Neural 7B\footcitetitleurl{IntelNeuralchat7bv31Hugging2023}, Mixstral 8x7B~\cite{jiangMistral7B2023}, Qwen1.5\footcitetitleurl{qwen1.5} and Llama3~\cite{llama3modelcard}.
Except for Mixtral 8x7B all of these models were since replaced by other more recent LLMs.
However, this does not decrease the validity of this throughput analysis.

While a throughput of 200 for the SSH connection is not impressive, it is sufficient for the purpose of our Chat AI service as the bottleneck is the capacity of the LLMs to respond to prompts.
If the model instances are scaled up by a factor of 10, it is possible that the infrastructure, especially the SSH connection, would need to be optimized or expanded to increase the capacity of the service.
Extending the HPC Proxy implementation to establish more than one SSH connection with the cluster and having it load balance requests would be feasible to overcome this limitation.

\subsection{User Adoption}

\begin{figure}[ht]
    \label{fig:usage1}
    \centering
    \includegraphics[width=0.75\textwidth]{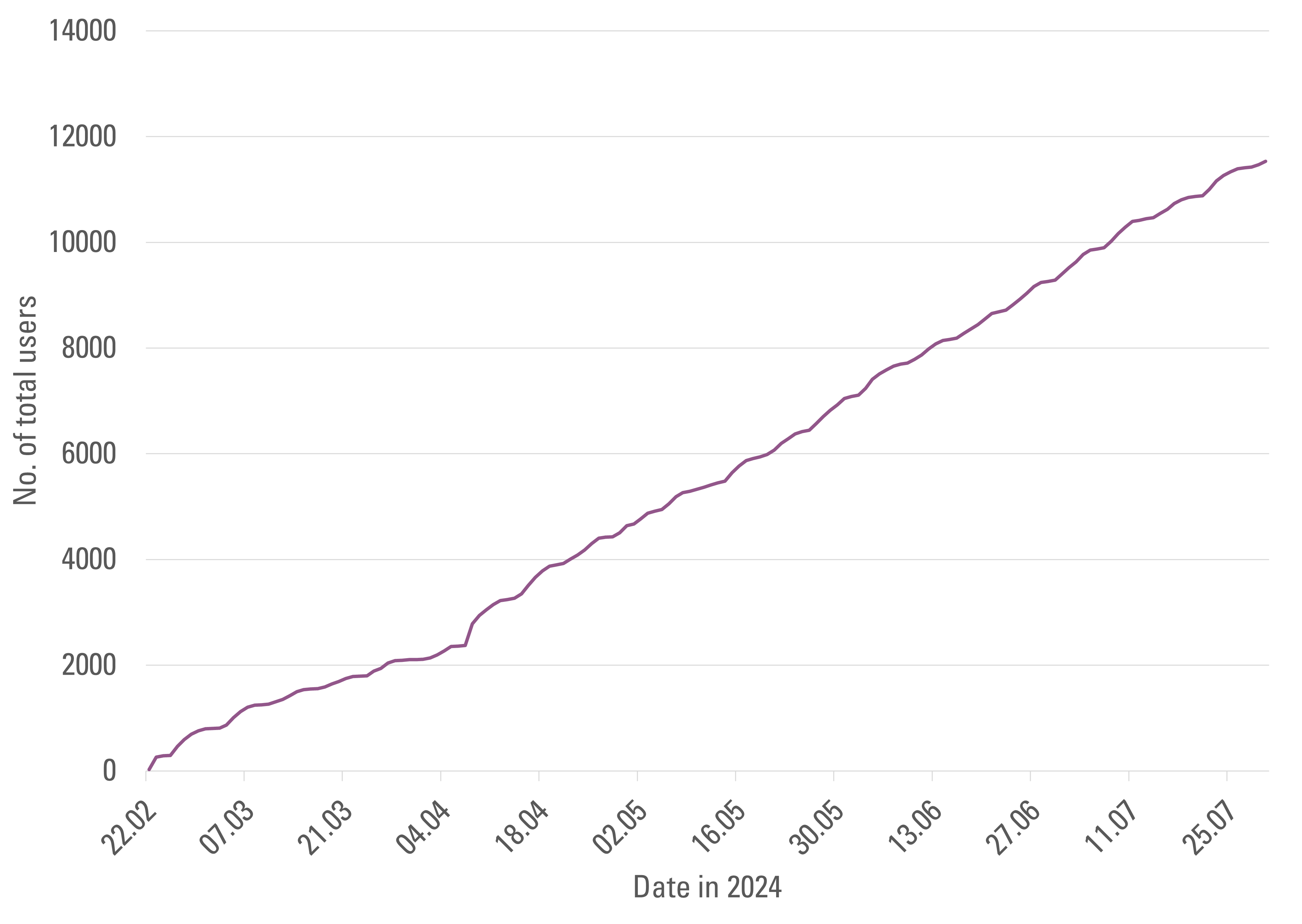}
    \caption[Total number of distinct users.]{Total number of distinct users from Feburary 22nd until July 30th 2024.
    The total number of users has grown consistently since its initial release, with a slight jump following a university-wide advertisement on April 8th.}
\end{figure}

Finally, we explore the user adoption and growth in popularity of Chat AI.
Since its initial release on February 22nd, 2024 to users of Academic Cloud\footcitetitleurl{AcademicCloud}, the popularity of the service has increased rapidly and consistently.
As shown in \Cref{fig:usage1}, in the first three months, over \(6\,000\) users were registered in the service, and by June~2024 this number increased to \(9\,000\).
This includes users from over 170 universities and over 150 research institutions with about 80 of the institutions being part of the Max-Plank-Gesellschaft (MPG)\footcitetitleurl{StartseiteMaxPlanckGesellschaft2024}.

\begin{figure}[ht]
    \label{fig:usage2}
    \centering
    \includegraphics[width=0.75\textwidth]{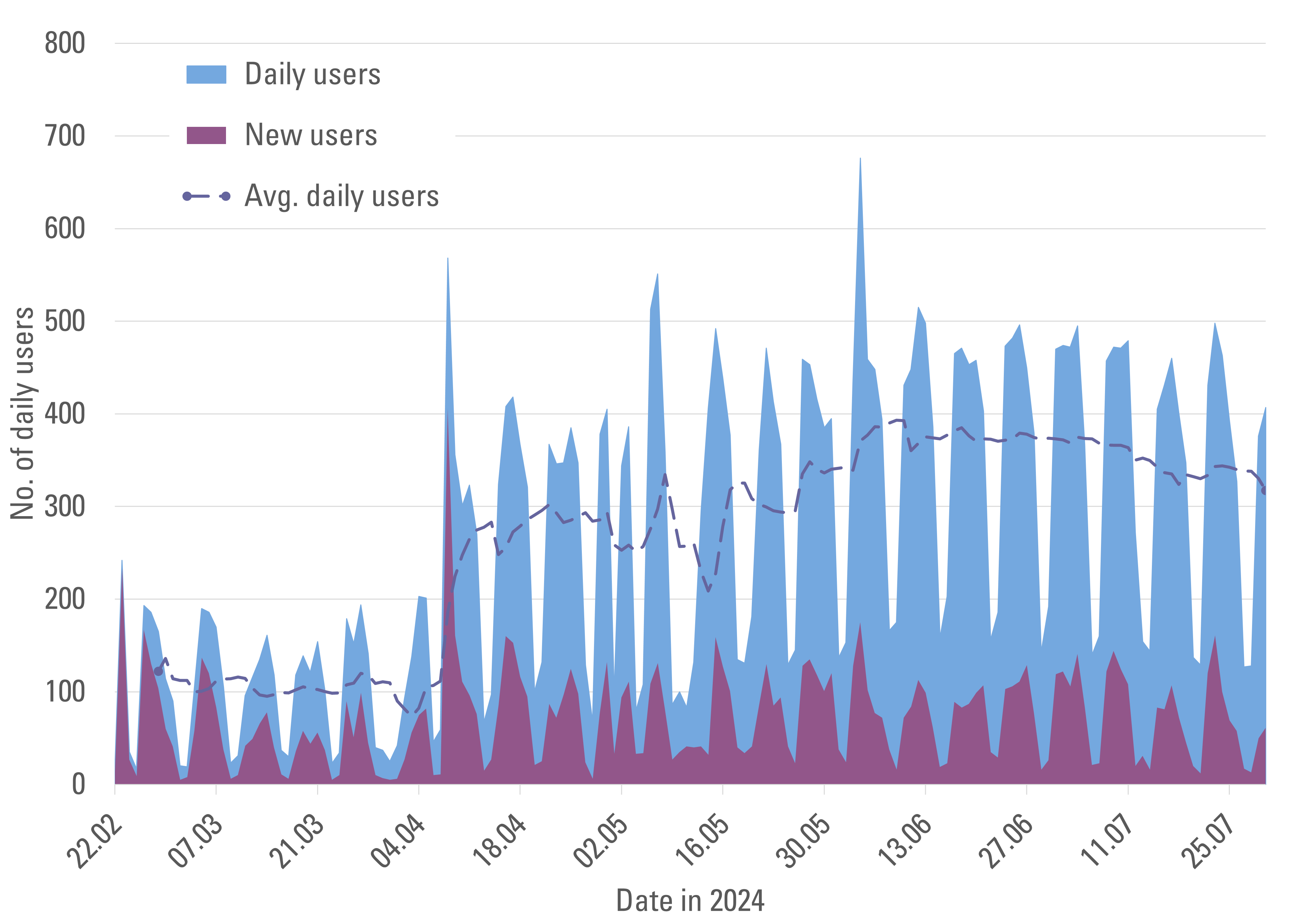}
    \caption[Daily Chat AI users.]{Daily Chat AI users from February 22nd until July 30th 2024.
    New users, are users that used the Chat AI service for the first time on a given day, while Daily users are returning users.}
\end{figure}

On average, about 400 to 500 users actively use Chat AI on a typical work day as can be seen in \Cref{fig:usage2}, of which approximately 100 are new to the service.
The drops are on the weekends and holidays, which indicates that our users mostly prefer to use Chat AI on weekdays rather than weekends, suggesting that Chat AI is mainly utilized in a professional setting.
A slight decrease in users can also be seen at the onset of the summer break in Germany in July.
In total, Chat AI has received and responded to more than \(350\,000\) messages as of 30.07.2024.
These numbers are expected to increase in the future as we continue to add features and improve the service.

\begin{figure}[ht]
    \label{fig:usage3}
    \centering
    \includegraphics[width=0.75\textwidth]{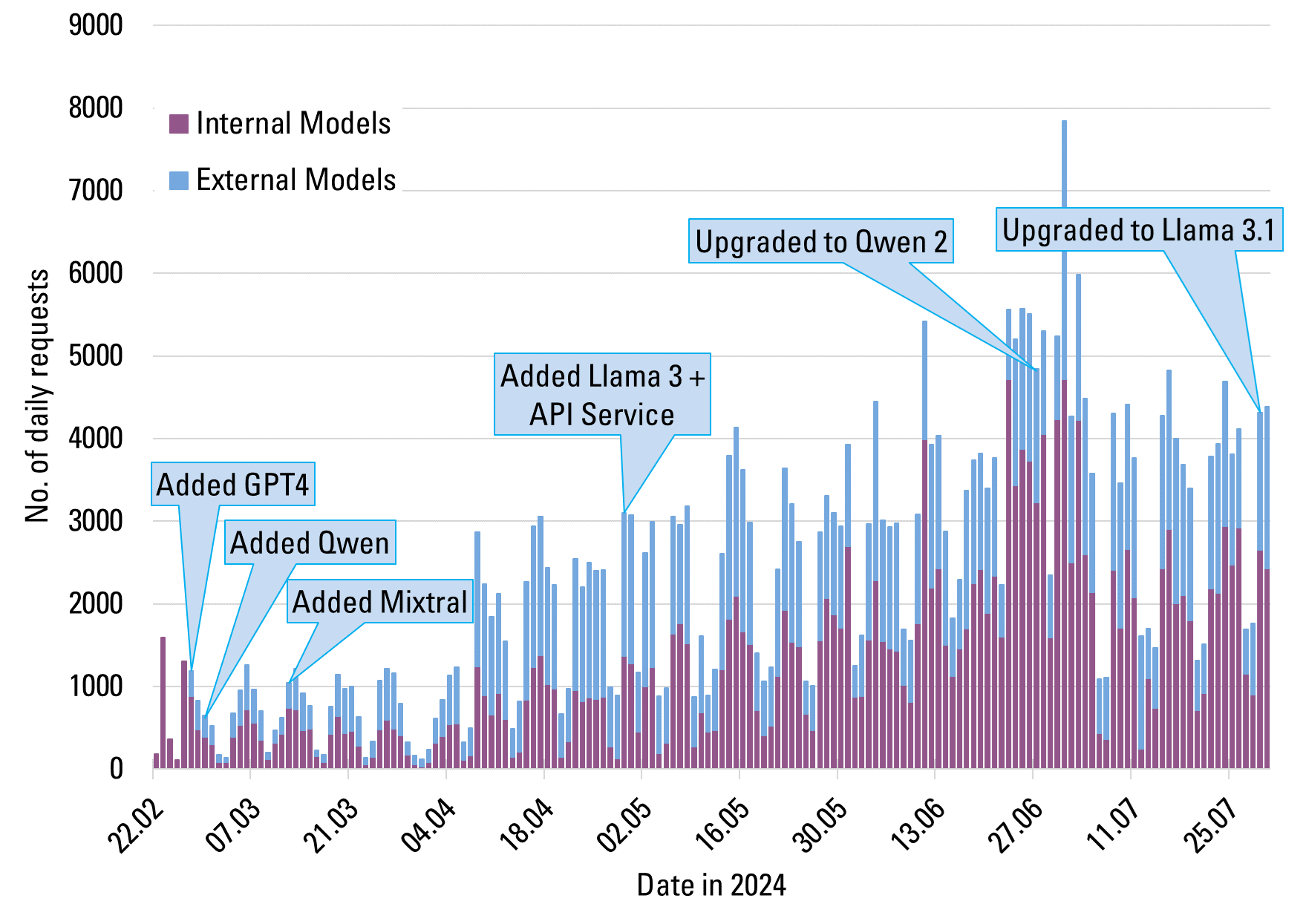}
    \caption[Total inference requests per day.]{Total inference requests per day from February 22nd until July 30th 2024.
    This shows the growth in popularity of Chat AI via the number of requests per day as well as when significant models were  added to the service.
    Moreover, each bar shows the amount of requests, which were handled via internal models, those hosted on our own infrastructure as opposed to external models, consisting of OpenAI's GPT3.5 and GPT4.}
\end{figure}

Over time, several features and models were added to Chat AI, which can be seen in the timeline given in \Cref{fig:usage3}.
The ability to interact with OpenAI's GPT-4 model was included shortly after the initial release of the service, followed by the addition of many state-of-the-art open source models in the subsequent weeks including Qwen and Mixtral as shown in \Cref{fig:usage3}.
Overall, four internal models are available through Chat AI at the time of writing including Meta's Llama3.1 8B and 70B, Alibaba Cloud's Qwen2 and Mistral's Mixtral 8x7B.

In May 2024, the redesign of the user interface was released, moving from our old gradio-based web interface to the new React and Vite-based interface.
With the redesign, several new functions and features were added such as custom system prompts.
A key feature that was introduced in response to popular demand was the possibility to store conversations onto local devices via an export and import functionality.
This enables the users to restore and resume previous conversations without compromising data privacy.

Additionally, we began offering OpenAI-compatible API access to our open source models, for researchers and students to integrate custom apps and run experiments with Chat AI.
The API access, which is provided on request, proved to be massively popular with over 100 API users and drastically increasing the number of requests and demand for the open source models.

Despite offering free access to GPT-4, a massively popular commercial model, many users still chose to interact with open source LLMs in Chat AI as can be seen in \Cref{fig:usage3}.
The major advantages, which could explain this demand is the data privacy, security, customization, inference speed, and more recently the ability to interact via the API.
This indicates that despite the apparent superiority of commercial models, there is sufficient demand for open source LLMs that it is possible to offer a competitive service without depending on commercial LLMs, by leveraging these advantages.
It also demonstrates the capability of existing HPC infrastructures to host computationally demanding web services using the proposed architecture at large scale.

\section{Discussion}\label{sec:discussion}

As the adoption of Chat AI continues to grow throughout universities and institutions in Germany, our contribution as well as its advantages are evident in enabling the ability to host a practical and secure web service on existing Slurm-based HPC clusters.
In this section we discuss the limitations of our approach and attempt to visualize the future outlook of our service. 

\subsection{Limitations}

Throughout our development efforts and the lifetime of the service, we encountered several problems and challenges, as well as, identified limitations of our design, some of which were resolved or could be resolved through additional development efforts but some remain open and would require significant changes.

These challenges most significantly include
\begin{itemize}
    \item reliability as in ensuring consistent availability of the service and recovering from failure states,
    \item automation to reduce the effort of admins to add new models to the service and
    \item scaling to zero the number of active instance for a given model when it is not required to release resources,
    \item additional security considerations regarding man-in-the-middle attacks and
    \item throughput limitations of our implementation.
\end{itemize}

\subsubsection{Reliability}

Some reliability issues are rooted in the architecture due to its many points of failure, such as the SSH connection from the web server to the HPC service node, which, if broken, can cause an outage of the entire service.
This was resolved by improving the implementation of the HPC Proxy to quickly and automatically restore the connection.

Another issue we encountered were the numerous failures of the LLM service on the HPC nodes, some of which were caused by bugs and incompatibilities of vLLM running on our GPU nodes, while some were due to the incapability of Slurm to handle cases where a node would fail, and, for example, required a reboot.
While these issues are not specific to our architecture, our scheduler script was not equipped to properly recover from these failure states without manual intervention.

Furthermore, in some cases during high-demand and multiple node failures, it was possible that Slurm could not find the required resources to start a new model instance, and a given model became unavailable in the meantime.
To mitigate these issues, we introduced some countermeasures in the scheduler script, but the reliability and stability still need further improvement.

While scalability was one of our main goals, which we attempted to solve through the automatic scaling of the model instances during high demand, this scaling might not occur fast enough to ensure a smooth and consistent service from the user perspective.
While the scheduler script is reasonably quick in detecting increased load, each new model still needs to perform a cold start before being able to serve traffic.
To start an instance of a model, the model has to first load into GPU memory, which can take as long as ten minutes for larger models on our hardware configuration.
When the service encounters many requests for a model and subsequently scales up its number of instances, the system may become overloaded before the new instances are ready to respond.
One possible solution is to preemptively scale up the instances when a large number of requests is expected to arrive based on observed usage patterns, e.g., based on time of day.

\subsubsection{Automation}

Many of the problems that a resource manager for a service paradigm has to solve and that are already solved through a platform such as Kubernetes had to be re-implemented in our solutions.
This meant at first more manual setup and configuration tasks, which were gradually being automated.
For example, the detection and handling of common error states, which are solved through health checks and automatic restarts had to be implemented from scratch.

The deployment of new LLM models on the HPC platform still requires manual effort to set up the service for the scheduler script and to create the respective routes in the API Gateway.
However, to facilitate the adoption of our solution, more automation tools should be implemented and added to the service.

\subsubsection{Scale to zero}

With our current design it is not possible for a service to properly handle scale to zero.
In an implementation of scale to zero, when no requests arrive for a given service for some time, it would reduce the number of desired active service jobs to zero.
If a request then arrived for a service that has scaled to zero, the request would need to be held in a queue until a single instance of a service job is ready to process the accumulated requests.
Depending on the service job, the cold start time for it might be multiple minutes, for example, to start a 70B LLM, which could result in timeouts for users.
Currently, with our architecture and its implementation, it would be possible to scale a model down to zero instances but there is no component that maintains a queue of requests while the model is loading.
While this could be implemented it would require significant changes, as the HPC Proxy and all components in the web server are also unaware of the status of the HPC platform and the available service instances, such that only the Cloud Interface Script can detect that a request cannot be answered until the model is ready.
Therefore, as this is a script and not an active component, the queue would need to be held in the web server, and it would need to query the availability status of the services by constantly probing the Cloud Interface Script.

Another option would be to scale models to zero on a fixed schedule.
An HPC cluster that is used to serve LLM inference requests by day, could be used to process regular Slurm jobs by night.
This could be implemented via two Cron jobs.
The first job would at the end of the work day create a backup of the last used configuration for the scheduler and provide an empty configuration instead.
With this the service jobs would expire and the resources could be used by regular Slurm jobs.
Then at the start of the work day, the second Cron job would trigger and replace the empty configuration with the backup created by the other job such that the service could resume regular operation during the day.
In order for the service jobs to be scheduled by Slurm without waiting for a backlog for other Slurm jobs, they would need to be submitted with higher priority.

\subsubsection{Encryption} 

In our current implementation, it would be possible for an attacker to perform man-in-the-middle attacks if that attacker manages to breach the Chat AI web interface as the user messages are not encrypted internally.
Protecting against this would require user clients to perform asymmetric encryption with decryption only happening on the HPC platform itself.
This could be implemented via API client libraries and provided to users for sensitive workloads.





\subsubsection{Throughput}

Our implementation delivers a throughput of about \(200\) requests per second but can be scaled up, for example, by deploying multiple instances of the HPC Proxy or the Chat AI web server and having Kong do load balancing across these multiple instances or by further optimizing the implementation of the code.
However, the component that is the most expensive to scale up for a data center employing our solution is the number of service jobs with LLMs that require multiple high-end GPUs.

Assuming a throughput rate of 8 requests per second for an LLM that runs on \(2\) high-end GPUs, \(25\) instances of this LLM would be required to max out the throughput of our HPC Proxy, which would require \(50\) high-end GPUs.
Via deployment of multiple instances of the HPC Proxy and Chat AI web server, a limit of \(3\,000\)\,RPS could be achieved, which would require \(750\) high-end GPUs.

We consider this as sufficient for most data centers, HPC centers and even national HPC centers.
For hyperscaler data centers that serve \(10\,000\) or even \(100\,000\)\,RPS, this solution is not sufficient but also not intended for.

\subsection{Future Work}
Chat AI and its underlying architecture have proven to be a viable solution for hosting HPC-based web services.
As the popularity of Chat AI has grown, so did the demand for more features, models and services.
There are a multitude of additional AI services that can be offered on our infrastructure and extend the scope of Chat AI.

As users are requesting to be able to run custom LLMs, which only a small subset of the overall user base would use, we plan to further investigate how we can implement scale to zero such that we can support a wide range of models without each of them constantly occupying valuable GPU resources.

Furthermore, with the release of vision language models (VLMs)~\cite{zhangVisionLanguageModelsVision2024} both in the open source~\cite{wang2023cogvlm, chen2024dragonfly} and the commercial space via GPT4o~\cite{openaiGPT4TechnicalReport2024} we are also looking to expand our implementation to support VLMs as well.
Moreover, as our architecture is not specific to LLMs we are also planning to implement support for other AI services that benefit from GPU acceleration via our HPC infrastructure such as audio transcription and text-to-speech services.

Finally, in order for our services to scale out more and to improve automation while relying on well-established software, we are considering switching out the underlying infrastructure for a Kubernetes-based platform.

\section{Conclusion}\label{sec:conclusion}

In this paper we presented a solution for securely hosting high-performance web services on Slurm-based HPC infrastructure and demonstrated its capabilities through our implementation of Chat AI, a high-performance responsive web service that provides access to state-of-the-art open source Large Language Models (LLMs).
Notably, our solution leverages existing Slurm installations with no special hardware or software requirements, creating a low barrier to entry for other institutions with similar infrastructure and security concerns.

Furthermore, we demonstrated the success and popularity of Chat AI through its user adoption throughout the academic community.
We inferred that the popularity of the service, especially the open source models, stems from the full control we provide to users over their data, conversations and messages, as well as the powerful web interface with features such as custom system prompts.
The OpenAI-compatible API access to the open source models proved to be popular and drastically increased the demand for these models.

We believe that the Chat AI service and its underlying architecture have the potential to democratize secure and private access to state-of-the-art LLMs, including open source models.
We hope that our contribution facilitates progress in the development of high-performance, secure web services, accelerating innovation and research in various fields.

\section{Code \& Acknowledgements}\label{sec:code}

\paragraph{Source Code}
We provide the web interface as a stand-alone software, including our Chat AI API service, which can be set up and run on a local machine or a web server at \url{https://github.com/gwdg/chat-ai}.
We also provide the software setup of our web server consisting of multiple containers, including the API gateway, the proxies, and monitoring tools at \url{https://github.com/gwdg/saia-hub}.
Finally, we provide the Cloud Interface Script that ForceCommand runs, along with the scheduler and tools to run the LLM servers at \url{https://github.com/gwdg/saia-hpc}.
\paragraph{Acknowledgements}
This work was supported by the Federal Ministry of Education and Research (BMBF), Germany under the AI service center KISSKI (grant no. 01IS22093A and 01IS22093B) as well as many colleagues at the GWDG, the joint data center of Max Planck Society for the Advancement of Science (MPG) and University of Göttingen.


\newpage
\printbibliography[heading=bibintoc, nottype=online, nottype=software, resetnumbers=true] 




\end{document}